\documentclass[10pt,letterpaper]{article}
\usepackage{opex3}
\usepackage{graphicx}
\usepackage{epstopdf} 

\begin{document}

\title{Broadband subwavelength focusing of light using a passive drain}

\author{Heeso Noh$^{1,2}$ Sebastien M. Popoff$^{1}$ and Hui Cao$^{1,*}$}

\address{$^{1}$ Department of Applied Physics, Yale University, New Haven, CT 06520, USA\\
$^{2}$ Department of Nano and Electronic Physics, Kookmin University, Seoul, Korea}

\email{* hui.cao@yale.edu} 



\begin{abstract}
Optical absorption is usually considered deleterious, something to avoid if at all possible. We propose a broadband nanoabsorber that completely eliminates the diffracting wave, resulting in a subwavelength enhancement of the field. Broadband operation is made possible by engineering the dispersion of the complex dielectric function. The local enhancement can be significantly improved compared to the standard plane wave illumination of a metallic nanoparticle. Our numerical simulation shows that an optical pulse as short as 6 fs can be focused to a 11 nm region. Not only the local field, but also its gradient are greatly enhanced, pointing to applications in ultrafast nonlinear spectroscopy, sensing and communication with deep-subwavelength resolution.     
\end{abstract} 



\section{Introduction}

Overcoming the diffraction limit and focusing waves on a deep-subwavelength scale have been widely pursued for light, microwave and acoustics. The tight focusing not only can strongly enhance linear and nonlinear interactions between waves and matter, but also greatly improve the spatial resolution in sensing, imaging and communication. It is well known that in a homogeneous medium, wave diffraction limits the size of a focal spot to half the wavelength. This limit can be understood in a simple case of a monochromatic scalar wave. Consider a spherical wave $e^{-ikr-i \omega t} / r$ converging to the origin $r = 0$, where $\omega$ is the angular frequency, $k = 2 \pi / \lambda$ and $\lambda$ is the wavelength. After passing the origin, the wave becomes diverging, of the form $e^{ikr-i \omega t} / r$. The total field is a superposition of the two, i.e, $e^{ikr-i \omega t} / r - e^{-ikr-i \omega t} / r \propto \sin(k r- \omega t)/r$, which is proportional to the imaginary part of Green's function~\cite{CarminatiOL2007} - the solution to the wave equation in a homogeneous medium.  The minus sign of the diverging wave causes a destructive interference of the incoming and outgoing waves, which removes the singularity at $r=0$. The focal spot has a width of $\pi / k = \lambda / 2$, the well-known diffraction limit. Therefore, diffraction is a consequence of destructive interference of the converging and diverging waves. 

One way of breaking the diffraction limit is to make the interference constructive by flipping the sign of outgoing wave, so that  the total field becomes $e^{ikr-i \omega t} / r + e^{-ikr-i \omega t} / r \propto \cos(k r- \omega t)/r $. This can be done by placing a resonant scatterer at the origin. Below the resonant frequency, the diffracted wave experiences a $\pi$ phase shift. For the visible light, the surface plasmon resonance of a metallic nanosphere might be used. The intrinsic absorption of metal, however, would weaken the interference effect.

An alternative is to utilize the absorption to eliminate the outgoing wave, so the total field is equal to the incoming one $e^{ikr-i \omega t} / r$. This can be realized by placing a perfect absorber at the origin. The field amplitude, albeit a factor of two smaller than in the first case, approaches infinity as $r \rightarrow 0$. Diffraction is thus overcome, as demonstrated by de Rosny and Fink in an acoustic wave experiment \cite{deRosnyPRL02}. They used a time-reversed sink to cancel the diffracted wave $e^{-ikr-i \omega t} / r$ and obtained a focal spot smaller than $\lambda/14$. This active drain can operate over a broad frequency range, but it requires prior knowledge of the incoming signal~\cite{kinslerPRA10} and a setup to generate the time reverse source at the focal spot. The idea of using a drain to remove the diverging wave has recently been extended to microwaves~\cite{ma2011evidence}.

In this paper, we propose a passive drain which is not driven externally and does not need prior knowledge of the temporal profile of the impinging wave.  This method is based on the coherent perfect absorber we recently developed \cite{ChongPRL10, WanSci11, NohPRL12}. Perfect absorption of coherent light by two- or three-dimensional metallic nanostructures can be achieved via critical coupling to surface plasmon resonances \cite{NohPRL12}. Although it was originally demonstrated for narrow-band operation, we will show here that perfect absorption can be realized over a continuous band of frequency by engineering the dispersion of the complex dielectric function. We demonstrate in numerical simulation that a 6.7 fs  pulse is focused from far field to a region of $\sim 13$ nm. The temporal shape of the pulse can be arbitrary; as long as the spatial wavefront matches the time-reversed radiation pattern of a surface plasmon resonance, the incident pulse will be completely absorbed. 

The perfect absorption of an ultrashort optical pulse results in strong localization of the electric field energy in space and time, which greatly enhances light-matter interactions, e.g., two-photon excitation of molecules. Not only the field amplitude but also the field gradient is enhanced by many orders of magnitude, when the incoming light is critically coupled to a multipole resonance of surface plasmon in the metallic nanoparticle. The steeply varying field can efficiently excite the dipole-forbidden transitions of molecules and probe the dark states. Hence, the broad-band perfect absorber may be useful for ultrafast nonlinear spectroscopy and sensing at the nanoscale. Furthermore, our scheme is general and applicable to other types of waves, e.g. acoustic wave, microwave, pointing to possible applications in shock wave lithotripsy, microwave communication with deep-subwavelength resolution \cite{MontaldoWRM04}. 

\section{Principle}

In principle, if a subwavelength object is placed at $r = 0$, the outgoing wave can undergo a phase shift and its amplitude may be reduced. This two effects are respectively due to the internal resonance and dissipation of the object. The destructive interference effect cancels the diverging part of the field only when the incoming and outgoing waves have the same amplitude and $\pi$ phase difference. Any deviation from these conditions leads to a subwavelength enhancement of the field. That is why the commonly used plane wave illumination of a metallic nanoparticle, that is dissipative and has resonance close to the optical excitation wavelength, produces a strong local field enhancement. In the introduction section we discussed two extreme scenarios to optimize the subwavelength field enhancement: adjusting only the amplitude or the phase of the outgoing wave. In this section, we will consider more general cases.

First, we show how to change the relative phase between the converging and diverging waves to make them interfere constructively. For simplicity, we limit our analysis to two-dimensional (2D) waves, the extension to 3D is straight-forward \cite{NohPRL12}. Let us consider a cylindrical wave propagates in the $x-y$ plane towards the origin; its magnetic field is along the $z$ axis. After converging at the origin, the wave diverges, and the total field can be written in the polar coordinates as $H_z(r, \theta) = H^{(2)}_m(k r) e^{i m \theta} + s H^{(1)}_m(k r) e^{i m \theta}$, where $H^{(1)}_m$ ($H^{(2)}_m$) is the $m$th-order Hankel function of the first (second) kind and represents an outgoing (incoming) wave. The complex number, $s$, is the relative amplitude of the outgoing wave; in the absence of dissipation the outgoing wave carries the same energy as the incoming wave, $|s|=1$. 
The phase of $s$ is determined by the solution to the Maxwell's equations, which gives $s=1$ in vacuum or lossless dielectric media. Although both $H^{(1)}_m(kr)$ and $H^{(2)}_m(kr)$ diverge at $r = 0$, the total field remains finite, because the diverging part of $H^{(1)}_m(kr)$ is canceled by that of $H^{(2)}_m(kr)$. This cancellation can be regarded as a consequence of destructive interference between the incoming and outgoing waves.

To switch to constructive interference, we must reverse the sign of $s$. One way is placing a cylinder of radius $R$ at $r=0$ to interact with the incident wave. Due to the cylindrical symmetry of the system, the angular momentum $m$ is conserved upon scattering, so is the polarization. Thus the total field outside the cylinder ($r > R$) has the same expression as before, but with a different value of $s$; inside the cylinder ($r < R$) $ H_z(r, \theta) = a J_m( n k r) e^{i m  \theta}$, where $J_m$ is the Bessel function of the first kind, $n = \sqrt{\varepsilon}$, $\varepsilon$ is the complex dielectric function of the cylinder, and $a$ is a normalization constant. By matching the fields at the cylinder surface ($r = R$), we solve for $s$ and get  
$s = {{[n J_m(n k R) H^{(2)}_m{'}( k R) - J_m'(n k R) H^{(2)}_m( k R)]}
/{[J_m ' (n k R) H^{(1)}_m( k R) - n J_m(n k R) H^{(1)}_m{'}(k R)]}}$,
where $J'$ ($H'$) is the first-order derivative of the Bessel (Hankel) function. 

\begin{figure}[htbp]
\centering
\includegraphics[width=5in]{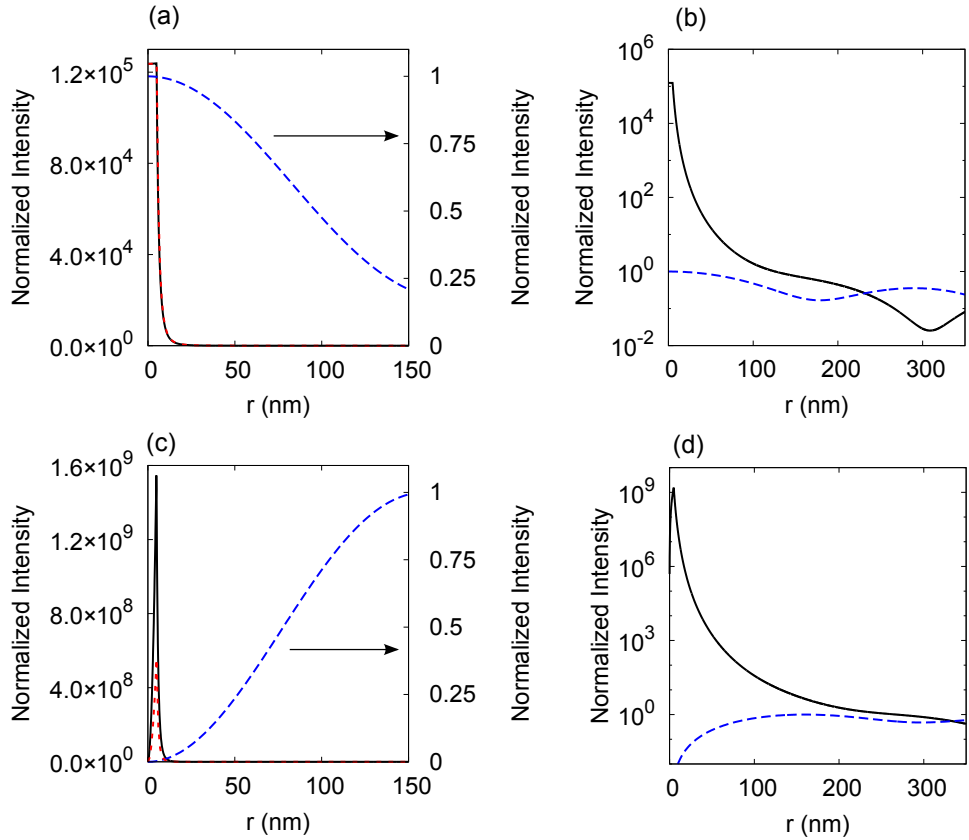}
\caption{Interference between a focusing wave ($\lambda =532$ nm) and its diffracted wave is made constructive ($s=-1$) by placing a lossless metallic nanoparticle ($R= 5.1$ nm; $\epsilon = -1.01$ for $m=1$; $\epsilon = -1.001$ for $m=2$) at the origin ($r=0$). Black solid curves (blue dashed curves) represent the radial distributions of electric field intensity with (without) the metallic scatterer for $m = 1$ (a,b) and $m = 2$ (c,d) in the linear scale (a,c) and the logarithmic scale (b,d). For comparison, the red dotted curves in (a,c) are the intensities when a plane wave is incident onto the metallic scatterer. All the intensities are normalized to the maximal values that can be obtained by focusing the same incident waves in vacuum without the particle. Maximal local field intensities are enhanced 5 orders of magnitude for $m = 1$ and 9 orders of magnitude for $m = 2$.}
\label{fig1}
\end{figure}

Setting $s = -1$ for a particular $m$, we find the complex $\varepsilon$ if $k$ and $R$ are fixed. Alternatively, we can fix $\varepsilon$ and solve for $k$ and $R$. Figure \ref{fig1} shows the dipole ($m=1$) and quadrupole ($m=2$) solutions for a cylinder of $R = 5.1$ nm. The wavelength $\lambda$ of light in the vacuum outside the cylinder is set at 532 nm. In both cases, the imaginary part of $\varepsilon$ ($\varepsilon_i$)  vanishes, the cylinder has no loss and $|s| = 1$. $\varepsilon$ is purely real and negative, representing a lossless metal. In the quasi-static limit, surface plasmon resonances occur at $\varepsilon=-1$. Below the resonant frequency (where $\varepsilon$ is slightly less than -1), the outgoing wave experiences an additional $\pi$ phase shift, and $s = -1$. The diverging parts of $H^{(1)}_m(kr)$ and $H^{(2)}_m(kr)$ no longer cancel each other giving rise to huge local field enhancements commonly observed with metallic nanoparticles.

The total field does not diverge due to the finite radius and reaches the maximal value at the surface of the cylinder ($r = R$). The smaller the $R$, the larger the maximal field magnitude. As shown in Fig. \ref{fig1}, the electric field intensity at $r=R$ is $10^5$ times higher than what can be obtained by focusing in vacuum without the cylinder for $m=1$. Nevertheless, a similar enhancement is obtained by exciting the same particle with a plane wave. When $kR \ll 1$, the input energy is predominantly coupled to the $m=1$ mode that has the lowest quality factor and whose radiation pattern has a better overlap with the plane wavefront. The enhancement originates from the same dipole resonance, thus the behavior is similar. However, for $m=2$, the field enhancement at $r=R$ is $1.5 \times 10^9$, which is about 3 times that with plane wave excitation. The quadrupole resonance ($m=2$) has stronger field enhancement, because of its lower radiative decay rate and higher quality factor. Since the plane wave excitation couples most of the input energy into the $m=1$ resonance that has a lower quality factor than the $m=2$ resonance, the field enhancement is weaker than that with a cylindrical wave of $m=2$. Excitation of higher-order ($m > 2$) resonances will lead to a further enhancement of local fields.

\begin{figure}[htbp]
\centering
\includegraphics[width=5in]{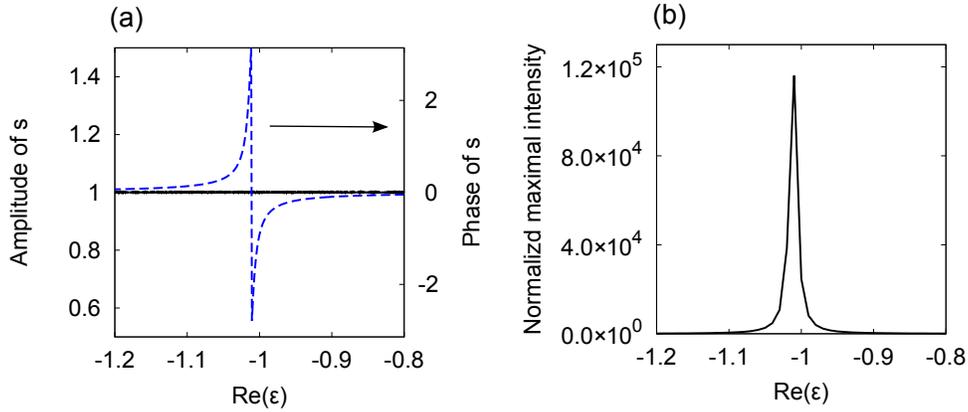}
\caption{(a) Relative amplitude (black solid line) and phase (blue dashed line) of the diffracted wave $s$ ($\lambda = 532$ nm, $m=1$) as a function of the real part of the dielectric constant $\varepsilon$ of the metallic cylinder ($R= 5.1$ nm). As soon as the value of Re[$\varepsilon$] deviates from that for $s = -1$, the phase of $s$ changes dramatically, breaking the condition for constructive interference of the scattered wave and the incident wave. (b) Maximal electric field intensity $I_e$ at the cylinder surface ($r= R$) as a function of the real part of $\varepsilon$. $I_e$ is normalized to the maximal field intensity produced by the same incident wave in the absence of the cylinder. When $s$ deviates from -1, $I_e$ drops quickly.}
\label{fig2}
\end{figure}

To describe the size of the focal spot, we use the effective diameter defined as $d_{eff} = 2 \int_0^{\infty}(|E_r|^2+|E_{\theta}|^2) r^2 dr / \int_0^{\infty}(|E_r|^2+|E_{\theta}|^2 r dr$. In Fig. 1, $d_{eff}$ = 18 nm for $m=1$, and $d_{eff}$ = 11 nm for $m=2$. These value are approximately $\lambda / 40$. Without the cylinder, the effective diameter of the focal spot is 212 nm, close to $\lambda/2$ - the diffraction limit. In fact, the focal spot size, which is slightly larger than the diameter of the cylinder (10.2 nm), can be further reduced by using a smaller cylinder. However, the value of $\varepsilon_r$ must be set correctly. 
As shown in Fig. \ref{fig2}(a), a slight deviation of $\varepsilon_r$ from the right value makes $s$ a complex number. The phase relation between the incoming and outgoing waves is modified, and they no longer interfere constructively. The total field can be expressed as the sum of a diverging term and a non-diverging term, $H_z(r, \theta) = 	H_{ND}(r,\theta) + (1-s)H_{D}(r,\theta)$, with $H_{ND}(r,\theta) = H^{(2)}_m(k r) e^{i m \theta} + H^{(1)}_m(k r) e^{i m \theta}$ and $H_D(r,\theta) = H^{(1)}_m(k r) e^{i m \theta}$. As $s$ deviates from 1, the maximal field intensity drops quickly [Fig. \ref{fig2}(b)].

\begin{figure}[htbp]
\centering
\includegraphics[width=5in]{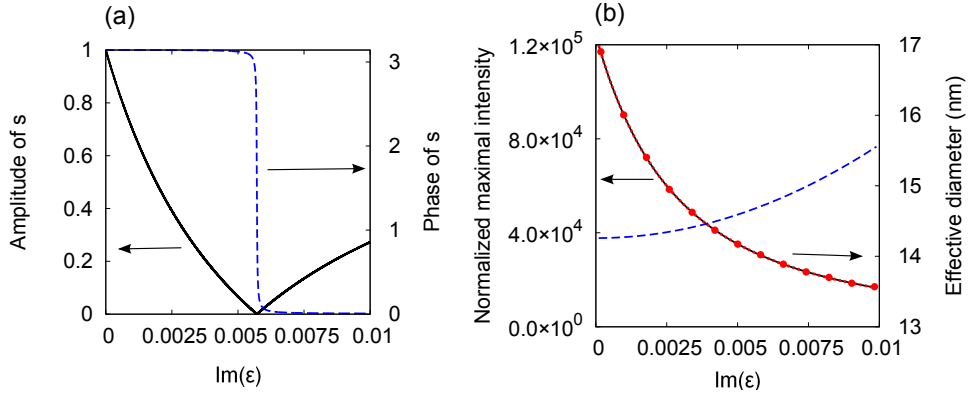}
\caption{(a) Relative amplitude (black solid line) and phase (blue dashed line) of the diffracted wave $s$ ($\lambda = 532$ nm, $m=1$) as a function of the imaginary part of the dielectric constant $\varepsilon$ of the metallic cylinder ($R= 5.1$ nm). The real part of $\varepsilon$ is fixed at -1.01. As the amplitude of $s$ reaches zero, its phase experiences a $\pi$ shift. (b) Maximal electric field intensity $I_e$ at the cylinder surface (black solid line) and the focal spot size defined by the effective diameter $d_{eff}$ (blue dashed line) as a function of the imaginary part of $\varepsilon$. $I_e$ is normalized to the maximal field intensity produced by the same incident wave in the absence of the cylinder. The red dotted curve represents the fit $I_e = \alpha \left|1-s\right|^2$, with $\alpha = 6.19 \times 10^4$. As the imaginary part of $\varepsilon$ increases from 0 to 0.01, $I_e$ decreases monotonically, while the effective spot size increases}.
\label{fig3}
\end{figure}

To achieve $s = -1$, the metallic nanocylinder must be lossless ($\varepsilon_i = 0$), which is difficult to realize at optical frequency. What happens if $\varepsilon_i \neq 0$? Figure \ref{fig3}(a) shows how the amplitude and phase of $s$ change as we fix $\varepsilon_r$ and gradually increase $\varepsilon_i$ for $m=1$. At small $\varepsilon_i$, the amplitude of $s$ drops, while the phase stays nearly constant. Hence, the absorption merely reduces the amplitude of the outgoing wave, but does not change its phase relation to the incoming one. The two waves still interfere constructively, albeit not as strongly as in the lossless case and the maximal field intensity decreases [Fig. \ref{fig3}(b)]. When $\varepsilon_i$ reaches a critical value $5.72 \times 10^{-3}$, $s = 0$, the outgoing wave vanishes, the incoming wave is completely absorbed via critical coupling to the nanocylinder cavity. With a further increase of the $\varepsilon_i$, the outgoing wave reappears, but with a $\pi$ phase shift. The interference between the incoming and outgoing waves becomes destructive, causing a further reduction of the maximal field intensity. The local field enhancement is determined by the diverging term of the total field, whose intensity is proportional to $|1-s|^2$, which fits well the maximal intensity in Fig. 3(b).  Since the destructive interference effect cancels the diverging part of the field only when $s$ is very close to $1$, the spot size is otherwise mainly governed by the huge local field enhancement close to the metallic surface. Nevertheless, as the imaginary part of $\varepsilon$ increases, the relative weight of the diverging term decreases and the effective diameter of the focal spot gradually increases in Fig. \ref{fig3}(b).

\section{Broadband Response}

The above study was performed at a single frequency. Is it possible to achieve subwavelength focusing for a broad-band pulse? To answer this question, let us consider the case $s = 0$; the solution can be easily extended to other values of $s$. $s=0$ corresponds to the perfect absorption of the incoming wave, which has been studied previously for narrow-band operation \cite{NohPRL12}. It utilizes critical coupling to the surface plasmon resonances in metallic nanostructures to achieve complete absorption. Since the surface plasmon resonances exist at discrete frequencies, the perfect absorption is narrow-band. Recently, an one-dimensional broad-band nearly perfect absorber was proposed and simulated by utilizing dispersive materials, e.g. the heavily doped silicon \cite{PuOE12}. 

\begin{figure}[htbp]
\centering
\includegraphics[width=5in]{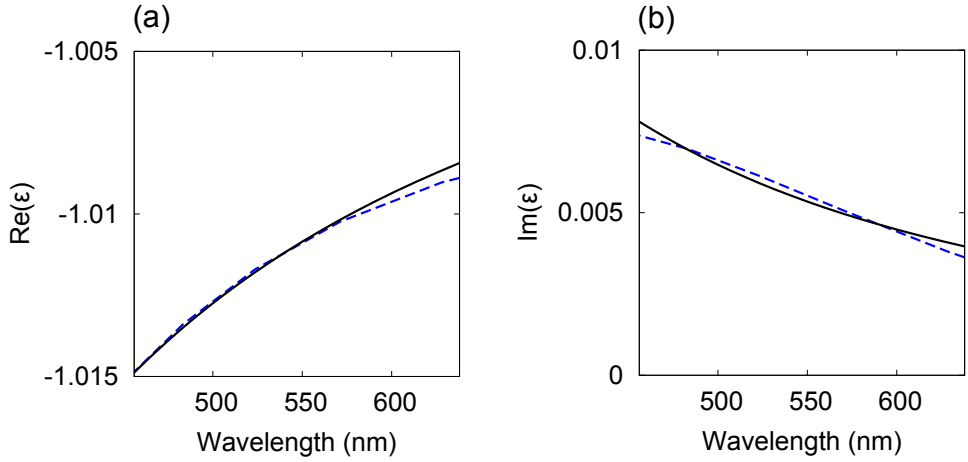}
\caption{Black solid curves in (a) and (b) are the real and imaginary parts of the dielectric function  $\varepsilon$ of a nanocylinder ($R = 5.1 $ nm) at $r=0$ for perfect absorption of a cylindrical wave ($m = 1$) at every wavelength in the range of 456 nm - 638 nm. The blue dashed curves represents the best fitting of complex dielectric function with the Lorentz model descried by Eq. \ref{eq1}. The parameters are given in the text.}
\label{fig4}
\end{figure}

Next we will show that by varying the complex dielectric function $\varepsilon$ with frequency, we can make the surface plasma resonate at all frequencies within a broad range and thus achieve the perfect absorption for a continuous band of frequency. As an example, we consider a nanocylinder of $R=5.1$ nm, and calculate the complex $\varepsilon$ for satisfying $s = 0$ at each wavelength $\lambda$ in the range of 456 nm to 638 nm. Figure \ref{fig4} plots $\varepsilon_r$ and $\varepsilon_i$ versus $k = 2 \pi / \lambda$ for $m = 1$. As $k$ increases, $\varepsilon_r$ decreases, and $\varepsilon_i$ increases. Such dispersion can be realized at the low frequency side of a material resonance.  We fit the dispersive $\varepsilon$ by the Lorentz model:
\begin{equation}
\varepsilon^{(f)}(\omega) = \varepsilon_{\inf}- {{\omega_p^2} \over {\omega^2 - \omega_0^2 + i \omega \Gamma_0}},
\label{eq1}
\end{equation}
where $\omega_0$ is the resonant frequency, $\Gamma_0$ is the damping constant, $\omega_p$ is the plasma frequency, $\varepsilon_{\inf}$ is the non-resonant high-frequency component of the dielectric function.

The dashed lines in Fig. \ref{fig4}(a,b) shows the best fit with $\omega_0 = 5.28\times 10^{15}$, $\omega_p = 6.67\times 10^{14}$, $\Gamma_0 = 5.85\times 10^{15}$ and $\varepsilon_{\inf} = -1.02+0.008i$. With such a dispersion relation, 
the cylinder can perfectly absorb the incoming waves $H^{(2)}_{m=1}(kr)$ across the wavelength range of 456 nm - 638 nm. Thus it can be considered as a ``white-light'' nanocavity, in analog to the macroscopic white-light cavities utilizing the negative dispersion of atoms \cite{WichtOC97}. 

\begin{figure}[htbp]
\centering
\includegraphics[width=5in]{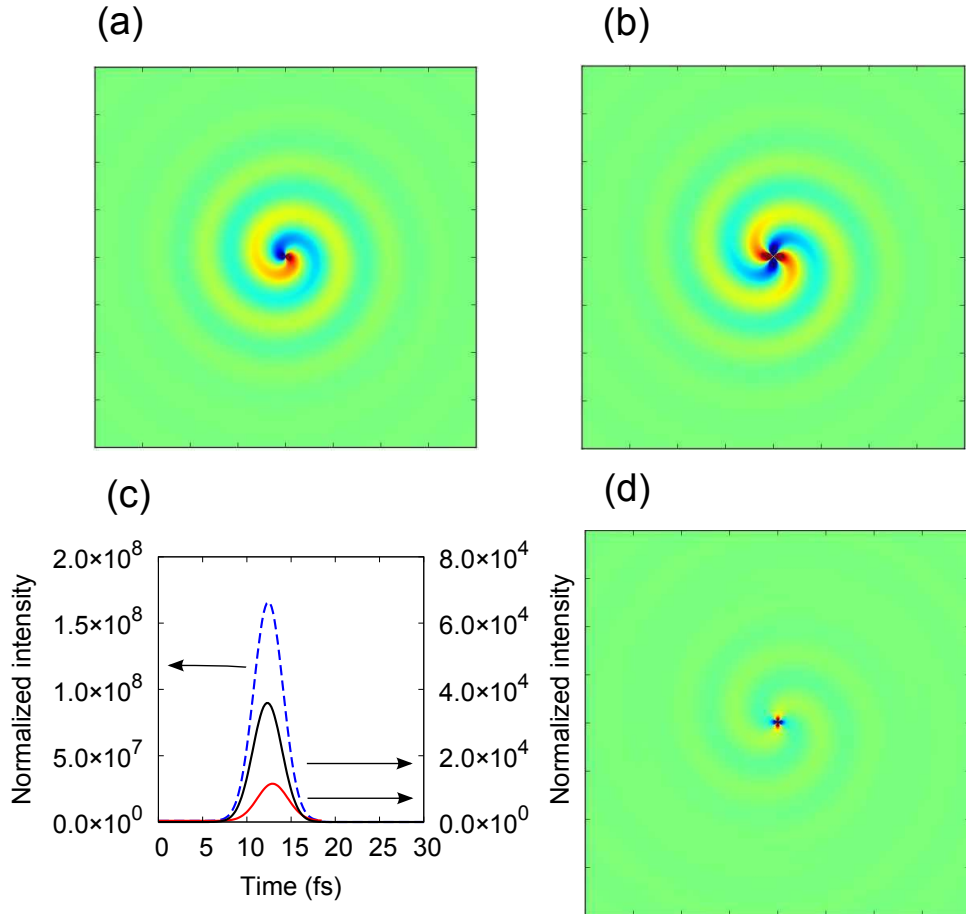}
\caption{(a,b) Single-frame excerpts from movies (Media 1, Media 2) of the spatial distribution of magnetic field $H_z(r, \theta)$ when a Gaussian pulse of width 6.7 fs impinges on a nanocylinder ($R = 5.1$ nm) at the origin ($r = 0$). The pulse has a cylindrical wavefront and an angular momentum $m = 1$ in (a), $m=2$ in (b). The pulse spectra is centered at $\lambda = 532$ nm with a FWHM of 180 nm. The dispersive dielectric function of the nanocylinder is chosen to reach a perfect absorption at all incident wavelengths. (c) Electric field intensity at the surface of the cylinder ($r=R$) as a function of time in the case of (a) - black solid line, and (b) - blue dashed line. The red curve represents the field intensity when the dielectric function is non-dispersive and perfect absorption is reached only at the center wavelength $\lambda = 532$ nm of the  incident pulse for $m=1$.  The intensity is normalized by the maximal intensity of the same incident pulse without the nanocylinder. (d) Single-frame excerpts from movies showing a chirped pulse (Media 3), whose spectrum and spatial wavefront are identical to the Gaussian pulse in (b), is perfectly absorbed by the nanocylinder with the same $\varepsilon(\lambda)$. The temporal profile of the pulse is described in the text.}
\label{fig5}
\end{figure}

The broadband operation enables a perfect absorption of a short pulse that is spread over the wavelength range of 456 nm - 638 nm. Figure \ref{fig5}(a) and (b) are movies (online) showing the temporal evolution of the field distribution when a Gaussian pulse of width 6.7 fs is impinging on the cylinder. The magnetic field of the incident pulse is $H_z = H_m^{(2)}(kr) e^{i m \theta} f(t)$, where $f(t) =e^{(t+r/c)^2/3.3^2}e^{-i\omega_c t}$, $t$ is in the unit of femtosecond, $\omega_c$ is $2\pi c /\lambda (\lambda=532\,{\rm nm})$ and $c$ is velocity of light. The incident wave has an angular momentum $m=1$ ($m=2$) in Fig. \ref{fig5}(a) [(b)], exciting the dipole (quadruple) resonance of the surface plasmon in the nanocylinder. In either case, the incoming pulse is completely absorbed by the cylinder, resulting in a nanoscale focal spot. Consequently, a strong local field builds up in the vicinity of the cylinder for a short period of time. Figure \ref{fig5}(b) plots the temporal evolution of the electric field intensity at the cylinder surface $I_e(r=R, t)$. It shows a peak of width 6.7 fs, equal to the incident pulse width. The peak height is 4 (8) orders of magnitude higher than the focused intensity of the same pulse without the cylinder for $m=1$ ($m=2$). As a comparison, we plot the temporal response of a non-dispersive particle that satisfies $s = 0$ for $m = 1$ at the central frequency of the pulse. The maximal amplitude of the pulse is 3.3 times weaker than that with the tailored dispersive particle. The focal spot size is $d$ = 18 nm, 11 nm for $m=1,2$. Therefore, the nanocylinder with an appropriate dispersion $\varepsilon(\lambda)$ enables the deep-subwavelength focusing of a broadband signal, that greatly enhances the local fields.

Note that with the same dispersive $\varepsilon$, optical pulses of arbitrary temporal shape can be completely absorbed, as long as all the spectral components are within the bandwidth of perfect absorption. We repeat the calculation with chirped pulses, and the incident wavefront remains cylindrical. Figure \ref{fig5}(d) is a movie of a pulse with the chirped temporal form $f(t) = e^{(t+r/c)^2/3.3^2}e^{-i(\omega_c-0.55t)t}$, $t$ in the unit of fs, and $m=2$ impinging on the nanocylinder. The pulse is focused to a $\lambda/44$  spot via perfect absorption of the diffracted wave.      

\begin{figure}[htbp]
\centering
\includegraphics[width=5in]{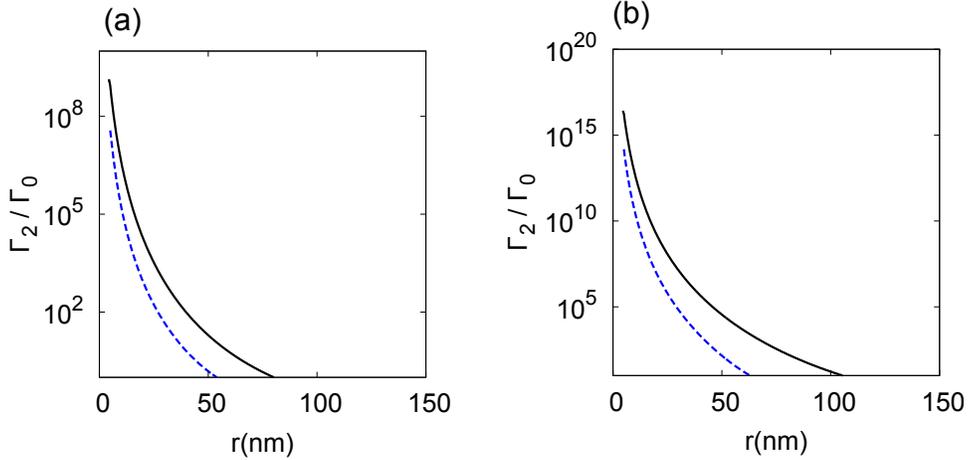}
\caption{Two-photon excitation rate $\Gamma_2$ as a function of the radial coordinate $r$ when a Gaussian pulse of width 6.7 fs is perfectly absorbed by a nanocylinder ($R=5.1$ nm) at $r = 0$. The incident pulse has the cylindrical wavefront of $m=1$ in (a), $m=2$ in (b) (black solid line). For comparison, the dashed blue curve represents the incident wave of a plane front. The pulse spectra and the dielectric function of the cylinder in (a) and (b) are identical to those in Fig. \ref{fig5}(a) and (b), respectively. $\Gamma_2$ is normalized by the maximal two-photon excitation rate $\Gamma_0$ of the same input pulse without the nanocylinder.}
\label{fig6}
\end{figure}

\section{Applications}

The subwavelength focusing of a short optical pulse greatly enhance the local field, which may be utilized to enhance the light matter interaction, especially the nonlinear one. As an example, we consider the two-photon excitation of molecules in the vicinity of a metallic nanocylinder ($R = 5.1$ nm) under the condition of broadband perfect absorption. The rate of two-photon excitation by a short pulse at a spatial position $(r, \theta)$ is $\Gamma_2 (r, \theta) \propto \int I_e^2 (r, \theta, t) dt $, where $I_e$ is the electric field intensity. Due to the cylindrical symmetry of the system and the excitation wave, $\Gamma_2$ is independent of $\theta$.  Figure \ref{fig6} plots the enhancement of $\Gamma_2$, i.e. the ratio $\Gamma_2(r) / \Gamma_{0}$, where $\Gamma_{0}$ is the maximal rate of local two-photon excitation by the same incident pulse in the absence of the nanocylinder. At the cylinder surface, $\Gamma_2 / \Gamma_{0} = 10^{9}$ for $m=1$, and $3 \times 10^{16}$ for $m = 2$. These enhancements are respectively $30$ and $300$ times higher than the ones obtained with the same short pulse having a plane wavefront ($\Gamma_2 / \Gamma_{0} = 3\times 10^7$ and $10^{14}$). Such enhancements occurs only locally ($r=R$). Moving away from the cylinder surface, $\Gamma_2$ decreases rapidly. The spatially-integrated rate, $\Gamma_{2t} = 2 \pi \int_{R}^{\infty} \Gamma_2(r) r dr$, is proportional to the total intensity of fluorescence from two-photon excitation. After normalizing it by that without the cylinder, we obtain the total enhancement factor, which is $9.7 \times 10^7$ for $m =1$, and $8.1 \times 10^{15}$ for $m = 2$. The enhancement factor can be further increased by using higher-order resonances ($m > 2$) or smaller scatterers ($R < 5$ nm).   

\begin{figure}[htbp]
\centering
\includegraphics[width=5in]{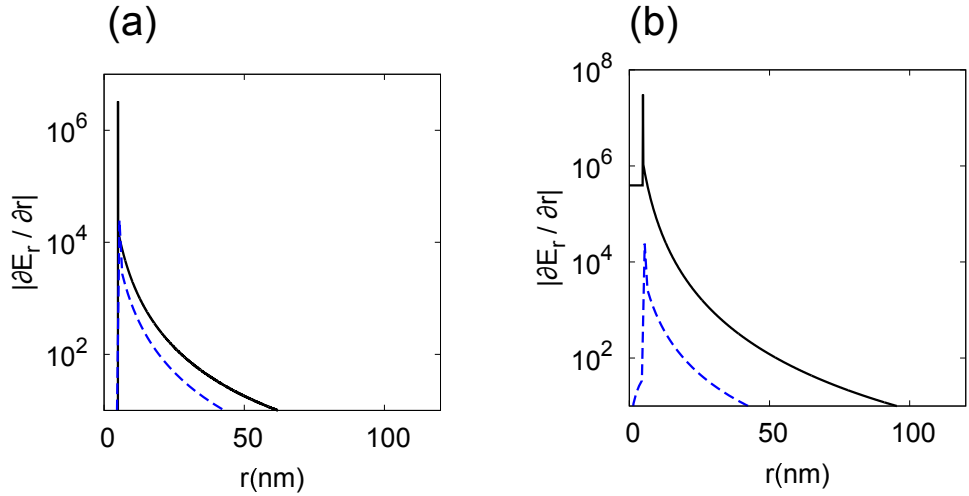}
\caption{Radial gradient of the electric field component in the radial direction $E_r$, in the case a cylindrical wave perfectly absorbed by a nanocylinder ($R=5.1$ nm) located at $r = 0$ (black solid line) and a plane wave excitation of the same cylinder (dashed blue line) at $\lambda = 532$ nm. $m=1$ in (a), $m=2$ in (b). $|\partial E_r / \partial r|$ is normalized by the maximal gradient without the nanocylinder.}
\label{fig7}
\end{figure}  

In addition to the field amplitude, the field gradient is also greatly enhanced. The steeply varying electric field can efficiently excite molecular multipole transitions that are barely excited, if at all, by far-field plane waves. For example, the rate of electric quadrupole transition is proportional to the spatial gradient of the electric field  \cite{YangCohenJPC11}. As an illustration, we calculate the electric field gradients in the case of perfect absorption at a single wavelength $\lambda = 532$ nm. For the TM-polarized light, the electric field has the radial component $E_r(r, \theta)$ and the azimuthal component $E_{\theta}(r, \theta)$. There are four gradients, among them the radial gradient of $E_r$ has the largest amplitude. Figure \ref{fig7} compares $|\partial E_r / \partial r|$ with a $R=5.1$ nm cylinder to that without for $m = 1$ [Fig. \ref{fig7}(a)] and $m = 2$ [Fig. \ref{fig7}(b)]. At the cylinder surface, $|\partial E_r / \partial r|$ reaches the maximal value, which is seven (six) orders of magnitude higher for $m = 2$ ($m = 1$) than the maximal value without the cylinder. By comparison, a plane wave excitation at the same wavelength gives rise to an enhancement of four orders of magnitudes lower in both cases. Such a dramatic enhancement of the field gradient will lead to an efficient excitation of the multiple transitions.    


\section{Conclusion}

In summary, we have proposed and simulated a broadband passive drain that can greatly improve the subwavelength focusing with a nanoparticle. It relies on the coherent perfect absorption to eliminate the outgoing (diverging) wave, whose interference with the incoming (converging) wave produces the diffraction pattern that limits the spatial resolution. The sink is not driven externally, and a broadband operation is made possible by engineering the dispersion of the complex dielectric function. Our numerical simulation shows that an optical pulse as short as 6 fs can be focused to a 11 nm region. This scheme can be easily extended to microwave and acoustic wave. The spectral dispersion for broadband response is reachable with carefully-designed metamaterials \cite{BelovPRE03, RahachouJOA07,FangXiNatMat06,DingHaoJAP10}. 

The broadband perfect nanoabsorber enables deep-subwavelength focusing of short optical pulses, which will greatly enhance nonlinear optical process. The spatial gradient of the field will also be enhanced dramatically, which may efficiently excite the dark states of molecules. Potential applications include ultrafast nonlinear spectroscopy, sensing and communication with deep-subwavelength resolution.  

We thank Mathis Fink, Geoffroy Lerosey and Seng Fatt Liew for useful discussions. 
This work is funded by the NSF Grants ECCS-1068642.

\end{document}